\documentclass[twocolumn,times,tighten]{aastex6}
\usepackage{amsmath,amstext}
\usepackage{multirow}
\usepackage[all]{hypcap} %Figure refs go figures (breaks deluxetables; use \capstartfalse \capstarttrue to fix it)
\usepackage{enumitem}
\setlist[enumerate]{leftmargin=*}

\shorttitle{Photo-stability of super-hydrogenated PAHs}
\shortauthors{M. Wolf et al.}

\begin{document}

\providecommand\natexlab[1]{#1}
\providecommand\JournalTitle[1]{#1}

\title{Photo-stability of super-hydrogenated PAHs determined by action spectroscopy experiments}
\author{M.~Wolf\altaffilmark{1}, H.~V.~Kiefer\altaffilmark{2}, J.~Langeland\altaffilmark{2}, L.~H.~Andersen\altaffilmark{2}, H.~Zettergren\altaffilmark{1}, H.~T.~Schmidt\altaffilmark{1}, H.~Cederquist\altaffilmark{1}, and~M.~H.~Stockett\altaffilmark{2}}
\altaffiltext{1}{Department of Physics, Stockholm University, 106 91 Stockholm, Sweden}
\altaffiltext{2}{Department of Physics and Astronomy, Aarhus University, Ny Munkegade 120, 8000 Aarhus C, Denmark}
\email{michael.wolf@fysik.su.se}

\begin{abstract}
We have investigated the photo-stability of pristine and super-hydrogenated pyrene cations (C$_{16}$H$_{10+m}^+, m = 0,6, \mathrm{\ or\ }16$) by means of gas-phase action spectroscopy. Optical absorption spectra and photo-induced dissociation mass spectra are presented. By measuring the yield of mass-selected photo-fragment ions as a function of laser pulse intensity, the number of photons (and hence the energy) needed for fragmentation of the carbon backbone was determined. Backbone fragmentation of pristine pyrene ions (C$_{16}$H$_{10}^+$) requires absorption of three photons of energy just below 3 eV, whereas super-hydrogenated hexahydropyrene (C$_{16}$H$_{16}^+$) must absorb two such photons and fully hydrogenated hexadecahydropyrene (C$_{16}$H$_{26}^+$) only a single photon. These results are consistent with previously reported dissociation energies for these ions. Our experiments clearly demonstrate that the increased heat capacity from the additional hydrogen atoms does not compensate for the weakening of the carbon backbone when pyrene is hydrogenated. In photodissociation regions, super-hydrogenated Polycyclic Aromatic Hydrocarbons (PAHs) have been proposed to serve as catalysts for H$_2$-formation. Our results indicate that carbon backbone fragmentation may be a serious competitor to H$_2$-formation at least for small hydrogenated PAHs like pyrene.
\end{abstract}

\maketitle

\section{Introduction}

Molecular hydrogen (H$_2$) is the smallest and most abundant molecule in the universe. Further, H$_2$ is the key to star formation and the starting point for the gas-phase chemistry of the interstellar medium (ISM) \citep{Tielens2005}. Polycyclic Aromatic Hydrocarbons (PAHs), along with fullerenes like C$_{60}$ \citep{Cami2010,Campbell2015}, may be among the largest molecules in the ISM. As a class, PAHs are widely believed to be responsible for the ubiquitous infrared emission bands observed throughout the ISM \citep{Tielens2008}. Despite their difference in size and complexity, the origin and fate of H$_2$ and PAHs may be closely intertwined in certain regions. Elevated H$_2$ formation rates have been observed in photodissociation regions (PDRs) with high PAH abundances \citep{Habart2003,Habart2004}. This has led to the suggestion  \citep{BauschlicherJr1998,Hirama2004} that PAHs may play a role as catalysts for H$_2$ formation in PDRs. It is, however, hard to understand how PAHs could survive in such harsh environments with strong UV-radiation. As a solution to this problem, \citet{Reitsma2014} recently suggested that PAH carbon backbones could be protected by the addition of hydrogen atoms and thus be able to catalyze H$_2$ formation in PDRs.

Super-hydrogenated PAHs (HPAHs), which contain additional H atoms beyond the native hydrogen already present in pristine PAHs, may, as indicated above, play an important role in molecular hydrogen formation \citep{Menella2012}. Formation of HPAHs should be efficient as the energy barriers for binding additional H atoms to PAHs are in most cases very low or even non-existent \citep{Rauls2008,Cazaux2016}. Highly hydrogenated species (up to and including fully saturated HPAHs) have indeed been produced in different laboratory experiments in which PAHs are bombarded with low-energy hydrogen or deuterium atoms \citep{Thrower2012,Boschman2012,Klaerke2013,Cazaux2016}, or through interaction with hydrogenated carbon surfaces \citep{Thrower2014}. Highly efficient H$_2$ emission from HPAHs has also been reported in infrared multi-photon dissociation \citep{Vala2009,Szczepanski2010} and UV matrix isolation spectroscopy experiments \citep{Fu2012}. This is in contrast to pristine PAHs, where single H-loss (although it is energetically disfavored \citep{Paris2014}) is much more common than H$_2$-loss at the excitation energies that are relevant in PDRs \citep{West2014,Chen2015}

While serving as nurseries for H$_2$ formation, HPAHs may also be protected by the additional H atoms. When excited by photo-absorption or through a collisions with energetic particles, HPAHs may relax by boiling off the weakly-bound excess H atoms, perhaps in the form of H$_2$. The existence of such a protective effect was inferred from the pioneering experiments by \citet{Reitsma2014}, where super-hydrogenated coronene cations (C$_{24}$H$_{12+m}^+; m=$ 0–-7) were found to lose fewer of their native H atoms than pristine coronene following core-electron excitation by soft x-rays. However, collision induced dissociation (CID) experiments by \citet{Gatchell2015} showed that super-hydrogenation of a smaller PAH, pyrene (C$_{16}$H$_{10+m}^+, m=0, 6, 16$), leads to a strong increase in the cross section for carbon backbone fragmentation. An earlier study of the photo-stability of small PAHs found that super-hydrogenation was associated with a greatly reduced barrier to single H-loss \citep{Jochims1999}. Although this work did not investigate H$_2$-formation or backbone fragmentation, it clearly indicated that single-H loss is an even stronger competitive channel for super-hydrogenated than for pristine PAHs. At a first glance this appear to speak in favor of the suggestion by \citet{Reitsma2014} as HPAH molecules would cool very efficiently by emission of the additional H-atoms.

Competition between hydrogen emission and backbone fragmentation may, however, depend on additional factors such as the size of the underlying PAH, the site(s) of hydrogenation, or the details of the excitation mechanisms. Is there, for example, a difference in the fragmentation of HPAHs excited with photons or in collisions with ions or atoms? In the case of x-ray excitation, highly electronically excited HPAH dications are formed as intermediaries and one may ask how the amount of internal excitation energy influences the fragmentation. Indeed it has been shown that the branching between H-loss and backbone fragmentation is highly sensitive to the internal excitation energy in other experiments on PAH dications \citep{Martin2012,Bredy2015}. In the CID experiments by \citet{Gatchell2015} and \citet{Wolf2016}, hydrogenated or pristine pyrene cations collided with He atoms at center-of-mass energies of 30--200~eV, simulating PAH-processing by supernova shocks \citep{Micelotta2010}. Collisions in this energy range activate nuclear motion directly in Rutherford-like scattering on the individual atoms in the molecule while they do not lead to significant electronic excitation or ionization. 

In this article, we present action spectroscopy measurements in which gas-phase pristine or hydrogenated pyrene cations (see Figure~\ref{fig: ELISA}) undergo fragmentation following the absorption of one or more optical photons with energies slightly below 3~eV. This mimics the conditions in the PDRs. We find that hydrogenation leads to larger, not smaller, rates of PAH carbon backbone fragmentation. This is clearly demonstrated by measurements of the backbone fragmentation yields as functions of laser power, from which we find the average number of absorbed photons needed to induce fragmentation. Pristine pyrene cations (C$_{16}$H$_{10}^+$) must absorb three photons for carbon backbone fragmentation, super-hydrogenated hexahydropyrene (C$_{16}$H$_{16}^+$) must absorb two, while the fully hydrogenated hexadecahydropyrene (C$_{16}$H$_{26}^+$) fragments after absorbing a single photon. Taken together with earlier collision experiments \citep{Gatchell2015,Wolf2016} and quantum chemical calculations \citep{Gatchell2015}, these results lead to clear conclusions. Regardless of the excitation mechanism, weakening of the carbon skeleton of small PAHs (by converting aromatic bonds to aliphatic ones) is more important than the cooling due to evaporation of additional H atoms in HPAHs. 

\section{Experiments}

\begin{figure}[ht!]
\epsscale{1.2}
\plotone{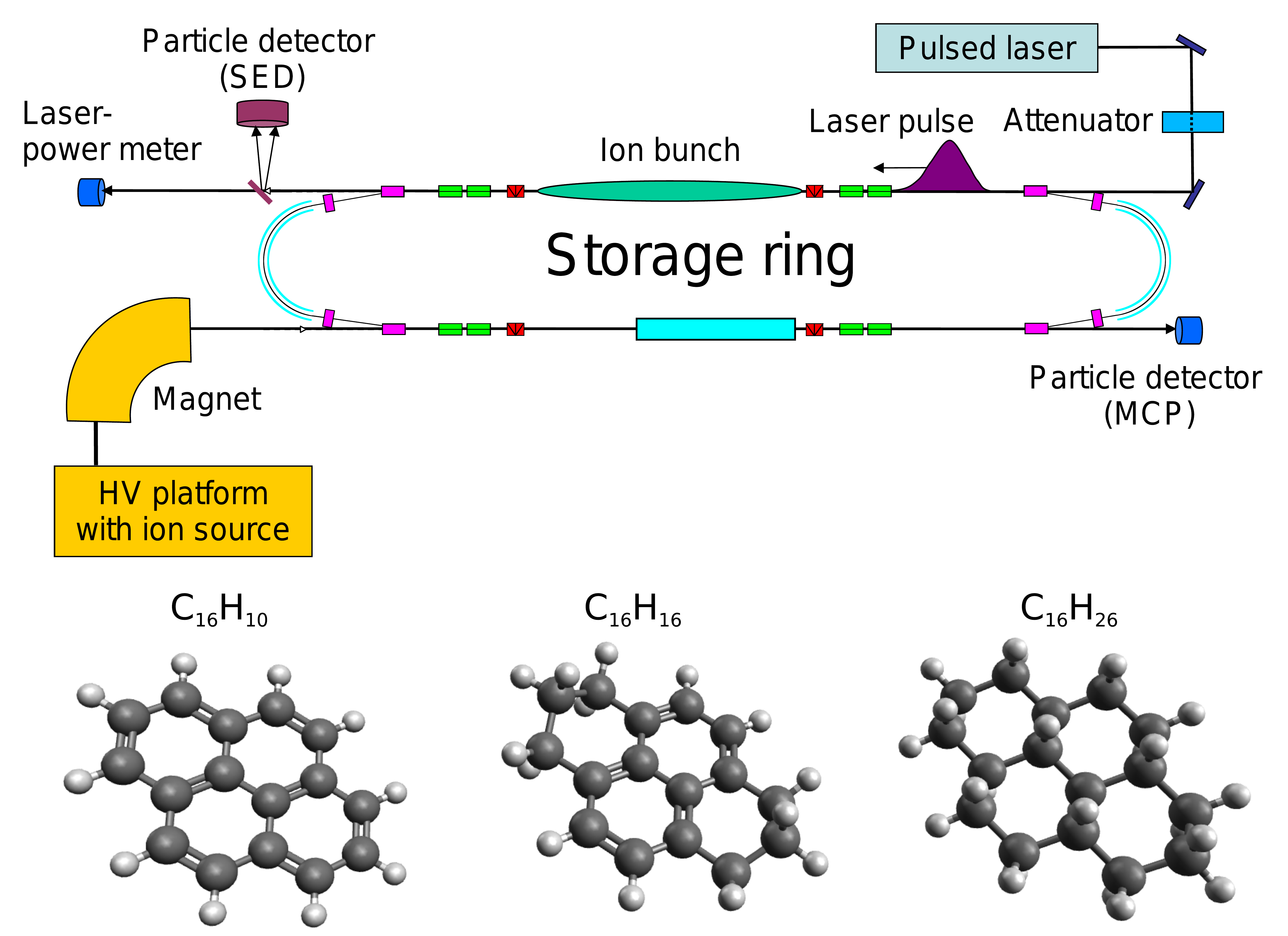}
\caption{Above: Schematic of the ELISA ion storage ring. Neutral fragments formed in the first few $\mu$s are detected with the Secondary Emission Detector (SED) (prompt action). Neutrals formed in the second straight section are detected with the Micro-Channel Plate (MCP) detector (delayed action). By switching the deflector voltages, daughter ions can be stored according to their mass/charge. \\ Below: Structures of C$_{16}$H$_{10+m}, m = 0,6, \mathrm{\ and\ }16$.\label{fig: ELISA}}
\end{figure}

Action spectroscopy experiments were conducted at the ELISA electrostatic ion storage ring at Aarhus University \citep{Moller1997,Andersen2002}. Pristine and hydrogenated pyrene radical cations were produced by electrospray ionization. Samples were purchased from Sigma-Aldrich and dissolved in dichloromethane, then mixed with a solution of silver nitrate in methanol \citep{Maziarz2005}, which was infused into a home-built electrospray ion source. The ions from the source were accumulated in a 22-pole ion trap containing He buffer gas for 50~ms prior to being gently extracted in bunches, and are considered to have equilibrated to room temperature. For the present molecules the corresponding internal energies of approximately 1--2~eV are significantly smaller than the photo-excitation energies. The ions of interest were selected by their mass to charge ratios using an electromagnet. Ion bunches were injected into the ELISA storage ring with kinetic ion energies of 22~keV and stored for 11~ms before being overlapped with a $\sim 4$~ns long laser pulse as shown in the upper part of Figure~\ref{fig: ELISA}. During the 4~ns irradiation time, the ions may absorb one or more photons sequentially, returning to the electronic ground state in between the individual photon-absorption events by converting the electronic excitation energy into internal vibrational energy. These hot ions may then dissociate on timescales ranging from a few ps to hundreds of $\mu$s. Neutral fragments formed while the ions are in one of the straight sections of the ring continue flying straight into one of two detectors. Neutral fragments formed in the first few microseconds after irradiation are detected with the Secondary Emission Detector (SED) \citep{Hanstorp1992}, constituting a ``prompt'' action signal \citep{Lammich2008}. Ions that survive at least one half turn (about 30~$\mu$s) or longer, and which decay in the second straight section of the ring, form neutral fragments which are detected with the Micro-Channel Plate (MCP) detector and contribute to the ``delayed'' action signal (see Figure~\ref{fig: ELISA}). The decay of the delayed action signal as a function of time (number of turns in ELISA) gives the dissociation lifetime. Three types of experiments were performed for each (C$_{16}$H$_{10+m}, m=0, 6, \text{ and } 16)$ molecule:

\begin{enumerate}

\item The total neutral fragment yields, both prompt and delayed, were recorded as functions of the excitation wavelength $\lambda$. This includes carbon backbone fragmentation as well as pure hydrogen losses. These action spectra are indirect measurements of the ion's electronic absorption spectrum. 
\item While irradiating at a fixed laser wavelength, the daughter ion mass spectrum was measured. This is accomplished by switching the voltages on all the electrostatic elements in the ring such that daughter ions (charged fragments) of a certain mass are stored instead of their parents \citep{Stochkel2008}. After one half turn (about 30~$\mu$s) the stored daughter ions are then dumped onto the MCP detector to measure the yield. This procedure is repeated as a function of secondary ion storage voltages to build the daughter ion mass spectrum. 
\item At a fixed wavelength, the laser power is varied using a continuously variable attenuator. From the resulting fragmentation yield curve, the average number of photons absorbed in the fragmentation process can be deduced. This measurement is performed both for the total prompt action and for mass-selected daughter ions.
\end{enumerate}

\section{Results and Discussion}

\subsection{Action Spectra}

\begin{figure}[ht!]
\epsscale{1.2}
\plotone{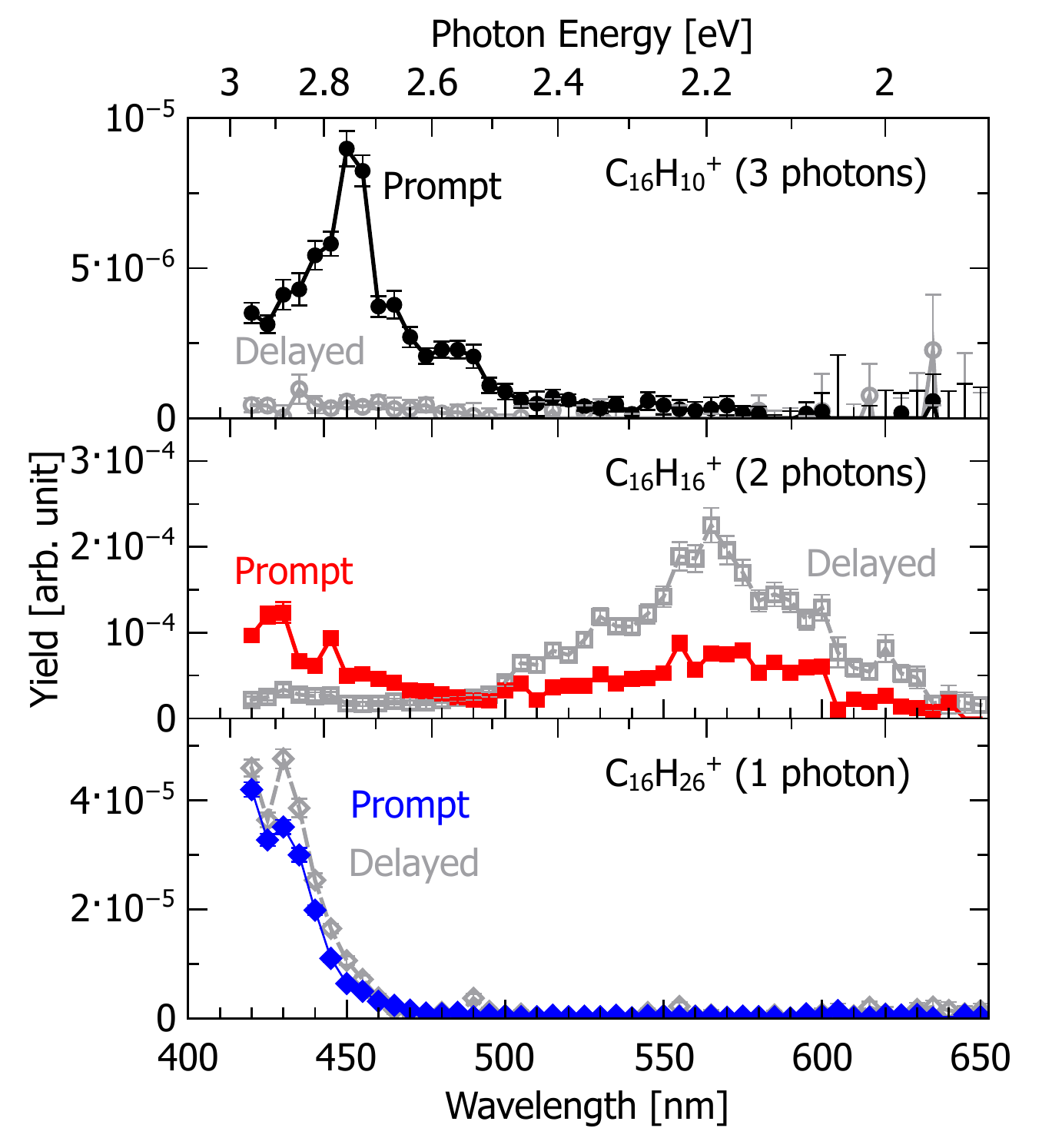}
\caption{Measured prompt (filled symbols) and delayed (open symbols) action spectra of C$_{16}$H$_{10+m}^+, m = 0,6, \mathrm{\ and\ }16$. The neutral fragment yield includes carbon backbone fragmentation as well as pure hydrogen losses. \label{fig: wavelengths}}
\end{figure}

In Figure~\ref{fig: wavelengths} we show the prompt and delayed action spectra of C$_{16}$H$_{10+m}^+ (m = 0,6, \mathrm{\ and\ }16$) measured at ELISA. The spectra have been corrected for the variation in the laser power across the spectral range and for the photon number dependence. This is achieved by dividing the background-corrected action signal at each wavelength by the number of photons per laser shot (the laser pulse energy divided by the photon energy) raised to the power of $n$, the number of photons absorbed in the dissociation process (determined in Section \ref{sec_powdep}, indicated in the Figure). It is non-trivial to extract the intrinsic absorption cross section from such a multi-photon dissociation action spectrum \citep{Wellman2015}, and clear interpretations of detailed band shapes are difficult.

The prompt action spectrum of pristine pyrene (C$_{16}$H$_{10}^+$, top panel) has a maximum at 450~nm. Delayed action with a dissociation lifetime of about 100~$\mu$s was observed, but the signal was quite weak. Previous gas-phase measurements of \textit{cold} pyrene cations in a supersonic expansion by \citet{Biennier2004} found an absorption band maximum at 436~nm. In a cold Ne matrix, this band is somewhat shifted to 440~nm \citep{Salama1993}. Multi-photon dissociation experiments performed by \citet{Useli-Bacchitta2010} using an ion trap observed the 436~nm band as well as a feature at 450~nm which was interpreted as a hot band. This is presumably the feature we observe in the present spectrum of room-temperature pyrene cations.

For C$_{16}$H$_{16}^+$ (middle panel), dissociation occurs with a lifetime of hundreds of $\mu$s. In the delayed action spectrum, a broad band with a maximum near 570~nm is observed. This band is also seen in the prompt action spectrum, where the dissociation yield is biased towards blue wavelengths. The absorption spectrum of C$_{16}$H$_{16}^+$ has not been reported previously, though it may be expected to resemble that of the naphthalene cation (C$_{10}$H$_8^+$) due to its similar $\pi$-orbital structure \citep{Halasinski2005}. Naphthalene has electronic transitions with origins near 670 and 455~nm \citep{Romanini1999, Pino1999}. Differences between prompt and delayed action spectra, whereby fragmentation induced by low-energy photons is observed on longer timescales than that induced by higher energy photons, are not uncommon in action spectroscopy experiments \citep{Lifshitz2002} and show the value of using an ion storage ring like ELISA to observe delayed fragmentation. 

The band maximum of fully hydrogenated C$_{16}$H$_{26}^+$ (bottom panel in Figure~\ref{fig: wavelengths}) appears to be below the lower limit of our laser's tuning range, but it is clear that the absorption is blue-shifted relative to that of pristine pyrene. In this case the prompt and delayed action (lifetime $\sim 50\ \mu$s) spectra are nearly identical.

\subsection{Daughter Ion Mass Spectrometry}

\begin{figure}[ht!]
\epsscale{1.2}
\plotone{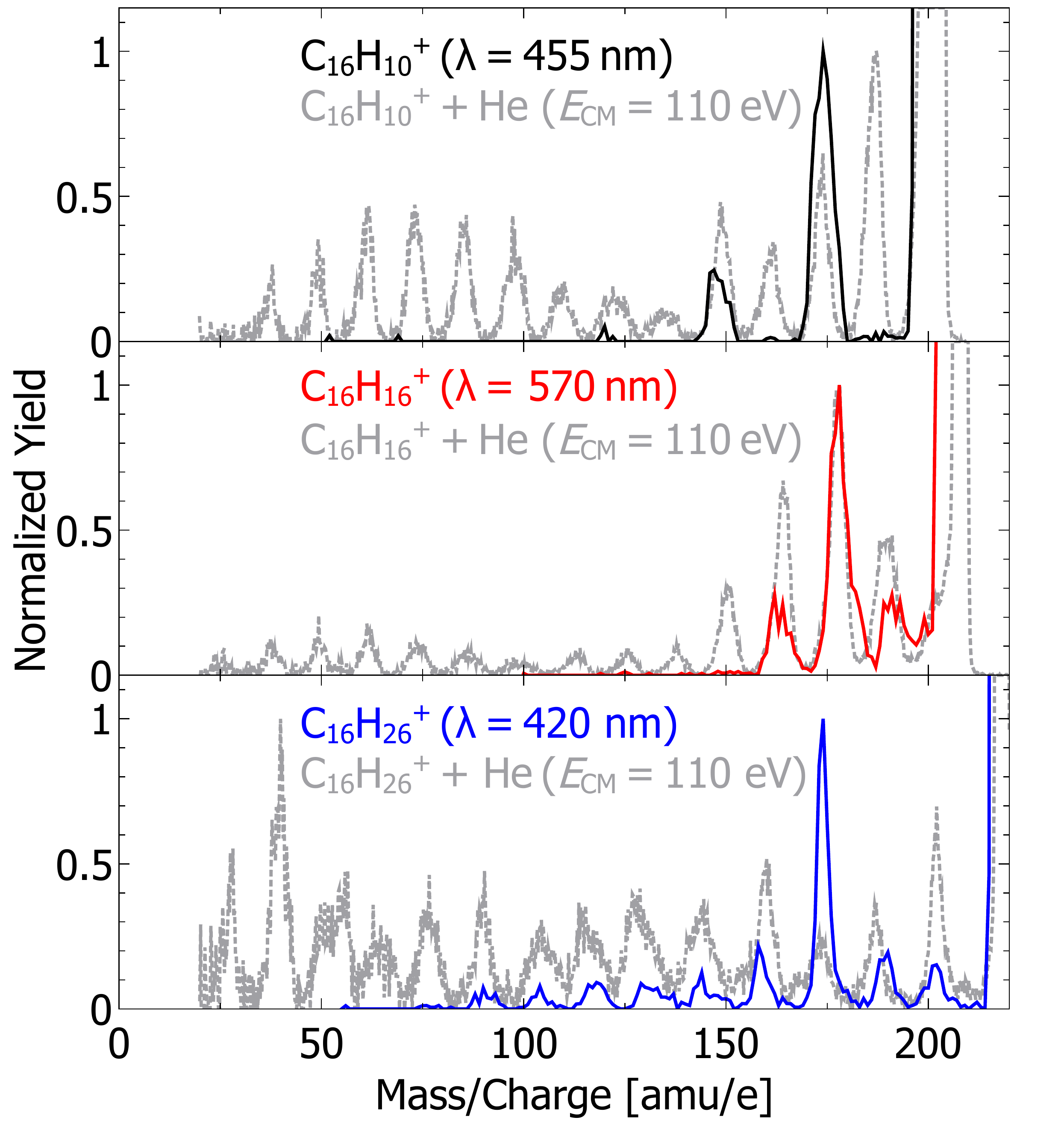}
\caption{Daughter ion mass spectra (full lines) of C$_{16}$H$_{10+m} (m = 0,6, \mathrm{\ and\ }16)$ when irradiated at the indicated wavelengths with full laser power. Dashed lines are CID mass spectra resulting from collisions with He at 110~eV center-of-mass energy, $E_\text{CM}$ \citep{Gatchell2015}.\label{fig: mass spec}}
\end{figure}

Fragmentation mass spectra, shown in Figure~\ref{fig: mass spec} (full lines), were recorded using excitation wavelengths near the absorption maxima in Figure~\ref{fig: wavelengths}. For C$_{16}$H$_{16}^+$, experiments were performed with both 420 and 570~nm excitation; the resulting mass spectra were found to be very similar. Carbon backbone fragmentation is clearly visible in all cases. Each peak in the mass spectrum corresponds to daughter ions with different numbers of carbon atoms; it is not possible to resolve individual hydrogenation states.  Also shown (dashed gray lines) are mass spectra for these same ions following collisions with He atoms at a center-of-mass energy $E_\text{CM} = 110$~eV previously measured at the DESIREE facility at Stockholm University \citep{Gatchell2015}. This is representative of the processing of PAHs by supernova shocks \citep{Micelotta2010}. Through nuclear stopping processes, such collisions deposit higher excitation energies (up to 40 eV \citep{Stockett2014b,Chen2014}) than the photon energies used at ELISA (below 3~eV), leading to more extensive fragmentation as can be seen in Figure~\ref{fig: mass spec}.

For pristine pyrene (C$_{16}$H$_{10}^+$, top panel of Figure~\ref{fig: mass spec}), peaks at masses corresponding to the loss of two and four carbon atoms are observed in the photo-induced dissociation (PID) mass spectrum at 176 and 160~amu. This is in agreement with the well-known tendency of small PAHs to decay via C$_2$H$_2$-emission \citep{Wacks1964,Ekern1998,Holm2010,Lawicki2011}. Notably absent in the PID spectrum for C$_{16}$H$_{10}^+$ are fragments which have lost a single C atom (\textit{i.e.} CH$_x$-loss). In CID such losses are due to non-statistical single carbon knockout \citep{Stockett2014,Stockett2015,Gatchell2014b} and clearly seen in the collision experiments. The mass spectra of C$_{16}$H$_{16}^+$ (middle panel of Figure~\ref{fig: mass spec}) are rather similar for the two experiments, although no fragments having lost four or more C atoms are seen in the PID data, presumably due to the lower excitation energy compared to CID. The presence of a CH$_x$-loss peak in the PID spectrum (around 192~amu) confirms earlier CID results which have shown statistical single carbon loss from HPAHs at collision energies below the knockout threshold \citep{Wolf2016}. For C$_{16}$H$_{26}^+$, the peak near 172~amu corresponding to the loss of three C atoms is much more prominent in the PID case than in the collision experiments, where all the fragment peaks are of similar intensity. 

\subsection{Laser Pulse Energy Dependence}
\label{sec_powdep}

\begin{figure}[ht!]
\epsscale{1.1}
\plotone{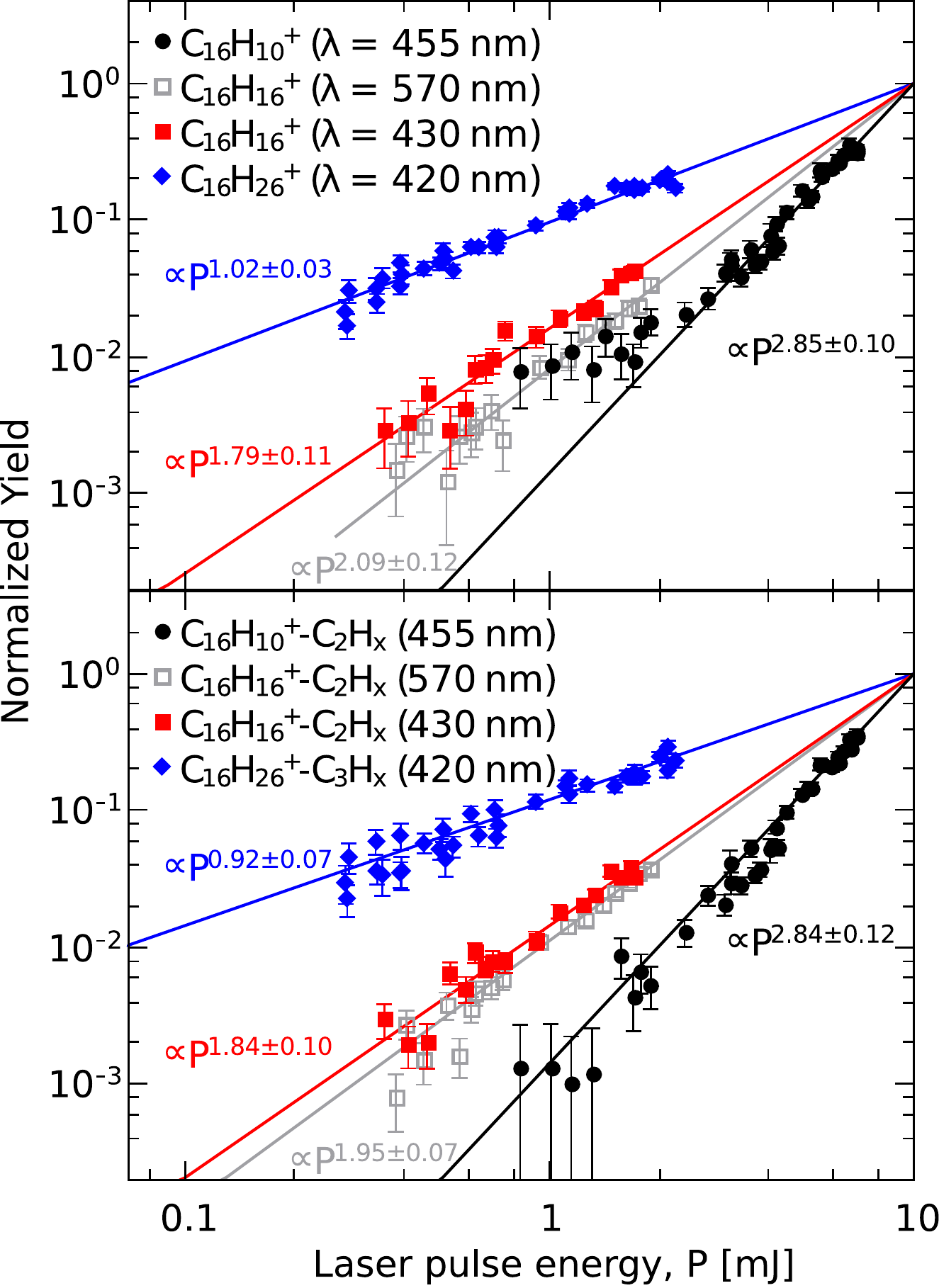}
\caption{Above: Laser pulse energy dependence of prompt action yields. The exponents found from power-law fits to the data give the average number of photons absorbed in the process. \\ Below: Laser pulse energy dependence of mass-selected daughter ion yields for the most abundant fragments (see Figure~\ref{fig: mass spec})}.
\label{fig: power dep} 
\end{figure}

The upper panel of Figure~\ref{fig: power dep} shows the total prompt neutral fragment yield as a function of the laser pulse energy. The excitation wavelengths are very similar and are chosen to be near the prompt action maxima (Figure~\ref{fig: wavelengths}). The yields follow power laws, where the exponents give the average number of photons absorbed \citep{Andersen2004}. For C$_{16}$H$_{16}^+$, we have performed such measurements both at 430~nm and at 570~nm, the maximum of the \textit{delayed} action spectrum. The photon number dependencies for 570~nm (open squares in Figure~\ref{fig: power dep}) were found to be nearly identical to those for 430~nm (closed squares).

Fits to the data give photon number dependencies of $2.85\pm0.10$ for C$_{16}$H$_{10}^+$, $1.79\pm0.11$ at 430~nm and $2.09\pm0.12$ at 570~nm for C$_{16}$H$_{16}^+$, and $1.02\pm0.03$ for C$_{16}$H$_{26}^+$. This is consistent with dissociation resulting mainly from three-, two-, and one-photon absorption events, respectively. The yield of the most abundant daughter ion for each molecule (see Figure~\ref{fig: mass spec}) as a function of laser pulse energy is shown in the lower panel of Figure~\ref{fig: power dep}. In each case, the power-law fit to the mass-selected daughter ion yield agrees with the corresponding total prompt action yield. 

Our experimental results are consistent with density functional theory calculations of the dissociation energies of these ions which have been reported by \citet{Gatchell2015}. Based on the laser pulse energy dependence, the total energy deposited in photo-absorption events leading to backbone fragmentation of C$_{16}$H$_{10+m}^+$ is 8.17~eV ($3\times 2.73$~eV), 5.77~eV ($2\times 2.88$~eV), and 2.95~eV for $m = 0,6, \mathrm{\ and\ } 16$, respectively. The calculated dissociation energies for the most abundant carbon-loss daughter ions are C$_2$H$_2$-loss with 6.30~eV, C$_2$H$_4$-loss with 3.88~eV, and C$_3$H$_6$-loss with 2.19~eV for $m = 0,6, \mathrm{\ and\ } 16$, respectively \citep{Gatchell2015}. While C$_2$H$_2$-loss is the backbone fragmentation channel with the lowest dissociation energy for pristine pyrene, C$_{16}$H$_{16}^+$ and C$_{16}$H$_{26}^+$ have CH$_3$-loss as their lowest dissociation-energy channels at 2.26 and 1.60~eV, respectively \citep{Gatchell2015}. Measurements of the CH$_x$-loss yields give photon number dependencies of $1.69\pm0.12$ and $0.95\pm0.15$ for C$_{16}$H$_{16}^+$ and C$_{16}$H$_{26}^+$, respectively, which are consistent with both the total prompt action and the other, more abundant, daughter ion channels measured for these two molecules. We note that, unlike the more abundant fragmentation channels, CH$_3$-loss requires an H-migration and may thus be slowed down by energy barriers.

In the upper panel of Figure~\ref{fig: power dep}, a deviation from power-law behavior is seen for the total fragmentation yield of C$_{16}$H$_{10}^+$ at the lower laser pulse energies. This could be due to a fragmentation channel with a lower activation energy (and hence power dependence), such as H-loss, which has a dissociation energy of 5.16~eV \citep{Gatchell2015}. This is below the excitation energy obtained by absorption of two photons ($2 \times 2.73=5.46$~eV) and together with the initial internal excitation energy of around 1~eV (room temperature value -- see the experimental section) this could give sufficient energy for H-loss on the microsecond timescale. Interestingly this kind of deviation is not seen in the measurement of the yield for the dominant fragmentation channel (C$_2$H$_2$-loss -- see the lower panel of Figure~\ref{fig: power dep}) as a function of laser pulse energy. The reason is, most likely, that the corresponding dissociation energy (6.3~eV) is too high to allow C$_2$H$_2$-loss on sufficiently short time scales.

\section{Summary}

Our results show that \textit{photo-induced} carbon backbone fragmentation is more likely for hydrogenated pyrene than for pristine pyrene. This is clearly demonstrated by the selected daughter ion laser energy dependence measurements. These experiments show that backbone fragmentation of pristine pyrene (C$_{16}$H$_{10}^+$) ions requires absorption of three photons of energy just below 3~eV, whereas super-hydrogenated hexahydropyrene (C$_{16}$H$_{16}^+$) must absorb two photons and the fully hydrogenated hexadecahydropyrene (C$_{16}$H$_{26}^+$) fragments after absorbing a single photon. 

In photodissociation regions (PDRs), where higher photon energies are common, it is likely that the carbon backbones of all three of these PAH ions would be photo-destroyed. What is clear from the present work is that pyrene is not protected by the additional hydrogen atoms. Carbon backbone fragmentation is thus likely to compete strongly with H$_2$ formation in PDRs, and this effect becomes more important with increasing degrees of hydrogenation. Our experiments provide important information for astrophysical models of PDRs, which have so far not taken into account the competition between backbone fragmentation and dehydrogenation of HPAHs \citep{Montillaud2013}. It remains an open question whether or not H$_2$ emission from HPAHs may occur prior to or in concert with the type of backbone fragmentation observed here. Fourier transform ion cyclotron resonance experiments such as those carried out on pristine pyrene cations \citep{West2014} could shed light on this issue. The present results suggest that it is unlikely that super-hydrogenation protects the carbon backbone for small PAHs, like pyrene, regardless of whether the excitation is by photons (at low energies) or by collisions with ions or atoms. An important next step is to conduct photo-stability studies of larger HPAHs like C$_{24}$H$_{12+m}$, to establish if hydrogenation effects are size dependent and perhaps protective for larger PAHs than pyrene.

\acknowledgments

This work was supported by the Villum Foundation and the Swedish Research Council (Contracts No. 621-2012-3660, No. 621-2014-4501, and No. 621-2015-04990). We acknowledge the COST action CM1204 XUV/X-ray Light and Fast Ions for Ultrafast Chemistry (XLIC). We thank Michael Gatchell, Linda Giacomozzi and Nathalie de Ruette at Stockholm University for providing valuable feedback on this manuscript and for other helpful discussions.

\end{document}